# Sentiment Analysis of the COVID-related r/Depression Posts

Zihan Chen

University of Ottawa

hchen154@uottawa.ca

Marina Sokolova

IBDA@Dalhousie University and University of Ottawa

sokolova@uottawa.ca

## Abstract
Reddit.com is a popular social media platform among young people. Reddit users share their stories to seek support from other users, especially during the Covid-19 pandemic. Messages posted on Reddit and their content have provided researchers with opportunity to analyze public concerns. In this study, we analyzed sentiments of COVID-related messages posted on r/Depression. Our study poses the following questions: a) What are the common topics that the Reddit users discuss? b) Can we use these topics to classify sentiments of the posts? c) What matters concern people more during the pandemic?

## 1. Introduction

We study sentiments in r/Depression posts related to COIVD-19. The Covid-19 pandemic has dramatically changed the way we have used to live. The pandemic has been causing significant devastations in economy, and health, *inter alia*. Mental health, especially, has become a growing concern due to employment terminations, income loss, family stress and other uncertainties. The pandemic disproportionally affected mental health of younger population. The Kaiser Family Foundation analysis of the Household Pulse Survey found that more than 56% of the young adults, 18- 24 years old, have reported the symptoms of depression during the Covid-19 pandemic. It is the highest rate among adult age categories, followed by 48.9% in the 25- 49 years old category (Panchal et al., 2021). People who lost their jobs during the pandemic and multi-generational families, with parents and children living together, report more symptoms of anxiety and depression than other population groups.

Sharing information on social media has played a critical role during the Covid-19 lockdowns and quarantines (Cauberghe et al, 2021). Reddit is a social media platform where people, mostly young adults, share their thoughts and experience. We collected the Reddit r/Depression posts to analyze the people's reaction to the Covid-19 outbreak. We use r/Depression as our text source. Our study poses the following questions: a) What are the common topics that the Reddit users discuss? b) Can we use these topics to classify sentiments of the posts? c) What matters concern people more during the pandemic?

This paper is structured as follows. Section 2 briefly overviews topic modelling and sentiment analysis studies of online texts related to COVID-19, Section 3 presents the methodology and framework of our study. In Section 4 we report empirical results of topic modelling and text classification. Section 5 presents sentiment analysis results and discusses the study limitation Conclusions and possible future work are discussed in Section 6. Appendix lists the toolkits used in this study.

## 2. Related work

In studies of the Covid-19 impact on the general public, sentiment analysis of the online posts helps to understand how the virus has affected personal lives and sense of well-being. This section lists papers closely related to our work, i.e., the authors apply topic modelling and sentiment analysis techniques to textual data from online sources. For the scoping review of early empirical studies on COVID-19 and social media we refer to Tsao et al (2021).

Topic modelling and sentiment analysis successfully use Machine Learning methods (Xie et al., 2020; Jelodar et al 2021). Some work focuses on microblogs, for instance Twitter and Weibo, whereas other turn their effort to longer texts, namely, mainstream media news and the Reddit posts (Sarlan et al. 2014; Balahur et al. 2010).

Research of sentiment analysis on Covid-19 news data was performed by Chakraborty and Bose (Chakraborty & Bose, 2020). They collected the news data from GDELT Project which contains online news from all over the world. In the research, they conducted sentiment and statistical analysis on the news data to uncover the relations between effect of Covid-19 and the sentiment of news. We, *au contraire*, work with messages posted by the Reddit users.

In the research targeting public opinion on Weibo, a Chinese microblog platform, Latent Dirichlet Allocation (LDA) has been used to discover the topics from posts on Weibo. Sentiment tagging based on lexicon-based approach was used to assign sentiment scores to the posts. They successfully explored the public responses to Covid-19 in China with both topics and sentiments (Xie et al., 2020). We, however, do not concentrate on one country and work with the Reddit social network available to users across the globe.

Murray et al. (2020) extracted symptoms (fever, sore throat) from personal narratives about Covid-19. The authors use the Reddit data. They used topics modelling and utilized the NRC sentiment lexicon approach to reveal the emotions in the reddit posts. The research reveals the sentiments across the first fourteen days of the Covid-19 infection. To the contrary, we use three tools (VADER, Textblob, SentiWordNet) to recognize expressed sentiments; also, we do not impose time restrictions on the expressed sentiments.

Jelodar et al (2020) analyse comments to the Reddit messages, posted on the Covid-19 related subreddit, not mental health. The authors used SentiStrenth to assign comments with different scores and then manually assigned sentiments based on the scores. We, however, used a fully automated sentiment labelling. Our work focuses on the COVID-related content of the messages posted on the Depression subreddit, i.e., we want to know topics that people talking about depression take the trouble to discuss during the pandemic.

# 3. Text Classification and Sentiment Analysis of the Reddit COVID-19 Messages

## 3.1 Goal of the study

In this study, we perform topic modelling and sentiment analysis of online depression discussions where people talk about Covid-19. Previous studies show that people with preexisting mental health conditions have experienced deterioration of the health conditions since the Covid-19 pandemic started (Alonzi et al., 2020). Our analysis focuses on contents on 'r/depression' subreddit from Reddit.com.

Sentiment analysis on the online public contents written by people experiencing depression under Covid-19 helps to understand the impact of Covid-19 on mental wellbeing of a large portion of population. Modelling the content topics of those posts allows researchers and policymakers to identify what people perceive as contributing factors to their mentally suffering during the Covid-19 pandemic and, consequently, give support to those people who are.

## 3.2 Summary of the study

Our study combines two tasks. The first task is to categorize unlabelled text data into categories related to the message content. We aim to identify a set of illustrative topics that can be used to represent the posts we collected. Since the posts are unlabelled and it is impossible to read large amount of text data and assign relevant topic to it manually, we utilize unsupervised statistical topic modelling to assign a relevant sentiment to each post. We apply Latent Dirichlet Allocation (LDA), a generative technique (Blei et al., 2003), to construct the message topics. Each message is then assigned with one central topic (work related, for example) and possibly several topics of lesser importance, with smaller contributions (%) than the central topic. Since we are only interested in the central topics, we discard the topics of less-importance in each text.

Afterwards, we utilize machine learning models to carry out text classifications to examine and evaluate the correctness of the topic labels. The second task is to perform sentiment analysis on the newly labelled data. We use the labelled data to train and evaluate our chosen models: traditional Machine Learning algorithms, i.e., Naive Bayes and Support Vector Machine (SVM), both applied on the tf-idf representation of the text data, and two state-of-art transformer-based algorithms, XLnet and Bert. Then, the model with the highest accuracy will be used to predict sentiments of the remaining unlabelled data. Analysis on relationship between topics and sentiments generated from task one and two will be carried out as final part of the project.

We set up three experiments to perform text and sentiment classifications. We chose the best model in sentiment classification to predict the data. Finally, we compared the performance of transformer-based models with traditional Machine Learning models in all experiments. We applied standard measure metrics to evaluate the performance of the models. We used Precision, Recall, F1-score and Accuracy in multi-class classification (Sokolova & Lapalme, 2009).

In a series of empirical experiments, we have selected the transformer-based model BERT (Devlin et al., 2018) as the prediction model after BERT has outperformed the other models in preliminary sentiment classification of the data. Bert is a multi-layer bidirectional Transformer encoder that uses pre-training on a large unlabelled corpus and fine-tuning on a labelled data for a specific task (Gonzales-Carvajal and Carrido-Merchan, 2021). Fig 1 presents the framework of our study.

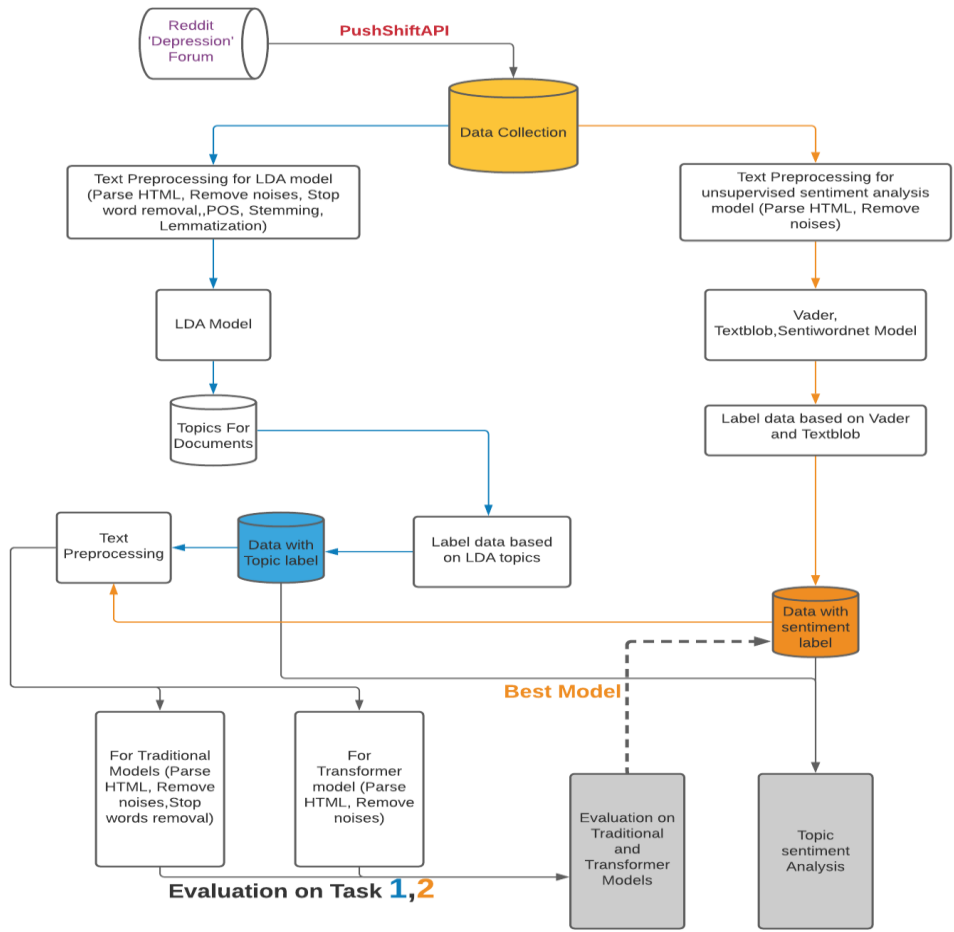

Figure 1: Methodology framework of the project

### 3.3 The r/Depression dataset

The Reddit online platform is considered as the 'the front page of the internet' and it has attracted millions of young adults and teenagers. According to (Auxier&Anderson, 2021), 36% of the people from 18 to 29 years old ever use Reddit; this is the largest portion among all the age groups. Reddit supports anonymity of the users that entices people to post their messages without undue reservations[1].Therefore, the posts could be subjective and make a rich resource for sentiment analysis tasks.

---

[1] redditinc.com/policies/content-policy

We extracted text posts from 'r/depression'[2]. Those subreddit posts convey self-expressed contextual aspects of depression and provide a richer context for sentiment analysis than more treatment-oriented posts from r/Anxiety and r/PTSD, the other mental health Reddit subcommunities (Park et al, 2018). Using PushshiftAPI (Baumgartner et al., 2020), we collected messages posted from Feb. 1, 2020, on which Covid cases started rising in North America, to Jan. 14, 2021. We set a filter parameter in the api to select only the attributes 'id', the body text of a post, 'title' and 'num_comments' from each api response. We initially retrieved 224,557 posts. Note that not all the posts refer to Covid-19. To appraise if a post relates to Covid-19, we built a list of covid-related keywords: "LOCKDOWN", "COVID-19", "PANDEMIC", "CORONAVIRUS" and "QUARANTINE. If a post has at least one such keyword and the number of text tokens in the post > 10, we save the post to our data collection list. We discard posts if they have 10 or less text tokens as they do not convey enough information for the statistical topic analysis. After we have removed the post duplicates, we have had 11,807 posts available for the topic modelling and sentiment analysis.

Online posts are affected by noise, i.e., any unplanned form different from the original text (Subramaniam et al., 2009). From the final data pool, we randomly selected 200 posts and manually analyzed their contents. The manual analysis has revealed dominant noise types that should be removed. We use the BeautifulSoup[3] to parse the texts and regular expression package, RE[4], to remove certain types of noise, e.g., special characters. We also adopt Contractions[5] to reconstruct text's contractions; for instance, restore "I cant do" to "I can not do". Further we work with the pre-processed, cleaned texts. Note that the resulting data contains stop words. On the next step, we transformed the textual data to the form that could be recognized by LDA model. More specifically, a list of tokens that contain meaningful and representative information of each post should be provided to a LDA model. We first ran the LDA model multiple times and removed the non-sequential words from the generated topics. We defined a list of non-sequential words, that we identified in the previous step, to be removed from the corpus. A sample of such words is provided in Figure 2.

```
non_keywords = ["year","month","thing","time","week", "day", "today","try", "make", "rant", "till", "aw", "hour","minute",
                "dont","didnt","havent","doesnt","wasnt","wont","isnt","wouldnt","kinda", "sorry"]
```

Figure 2: non-sequential keywords

Further, we tokenized our posts and obtained corresponding Part-Of-Speech (POS) tag for every word that does not appear on the stopword list. On the next step, we have selected words with the POS tags of adjective, noun, or verb, to represent the posts since such parts of speech provide most information in the text. Next, lemmatization and stemming were performed to group words based on single word from their inflected forms to make the denser data.

---

[2] https://www.reddit.com/r/depression
[3] https://www.crummy.com/software/BeautifulSoup/bs4/doc/
[4] https://docs.python.org/3/library/re.html
[5] https://github.com/kootenpv/contractions

# 4. Topic Modelling

## 4.1. The Optimal LDA model selection

There is no golden standard to select an LDA model that better fits a given task and the corresponding data. We aim to choose a model that provides an optimal solution for our study. In this, we use a two-criteria approach: criteria # 1- the topic coherence, criteria # 2 – the model's usability for text classification. To compute the topic coherence, we apply the four-stage topic coherence framework: segmentation, probability calculation, confirmation measure, aggregation (Röder et al, 2015). through from Gensim[6]. We evaluate the topic coherence with $C_v$, due to the measure's reliable performance on short texts (Syed & Spruit, 2017). We build several $LDA(i)$ models, where $i$ – the number of topics, then we compute $C_v$ – shown in Fig 3. We change only the number of topics and keep all other parameters as control variables in the evaluation.

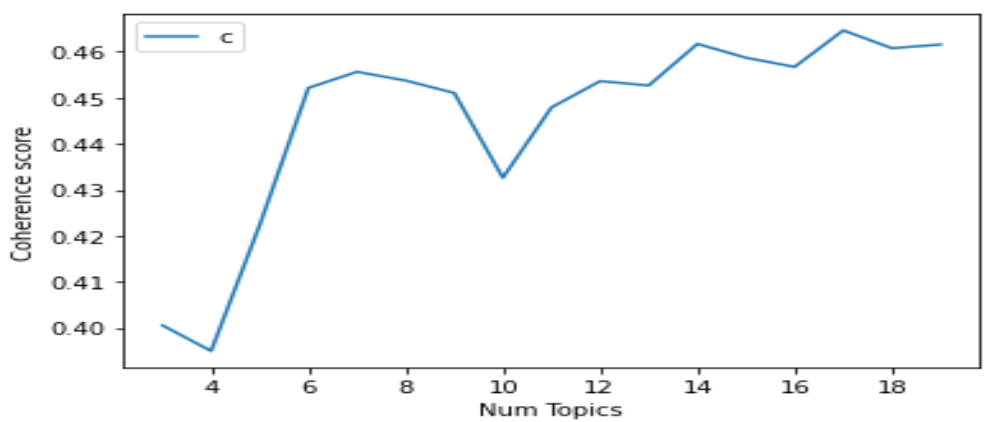

Figure 3: coherence values of models with different num_topics

The higher coherence values are obtained with 7, 12, …, 18 topics. For each topic number we calculate the baseline, naïve, multi-class classification accuracy, i.e., when the posts are equally assigned to the number of classes equal to the number of topics. For example, $LDA(15)$ has the coherence value 0.457, provides for 15 classes in multi-class classification, and achieves the naïve multi-class classification accuracy 6.67 %. Table 1 reports the results.

|  | 7 | 12 | 13 | 14 | 15 | 16 | 17 | 18 |
|---|---|---|---|---|---|---|---|---|
| $C_v$ | 0.454 | 0.453 | 0.451 | 0.458 | 0.457 | 0.454 | 0.463 | 0.458 |
| Baseline accuracy | 14.29 % | 8.33% | 7.69% | 7.14% | 6.67 % | 6.25% | 5.88% | 5.56% |

Table 1: The topic numbers that achieve the highest coherence values and corresponding naïve multi-class classification accuracy.

---

[6] https://radimrehurek.com/gensim/models/coherencemodel.html

LDA(7) becomes the candidate model as 7 topics provide for the highest baseline classification accuracy 14.29 %, LDA(7) archives the coherence value 0.454. In text classification, the high-quality topics become the driving factor of classification decisions instead of the most coherent topics (Stevens et al., 2012). By manually studying those topics, we found that most of the terms generated by LDA(7) produces relatively more interpretable as shown in the first visualization in Section 4.2 and Table 2. These terms appear in the forms of noun, verb, and adjective, which allow us to understand the contents easily (Sec. 3.3). LDA(7) becomes our model of the Reddit data.

Table 2: Topics generated by LDA(7)

|         | Keywords * Weights | Content focus |
|---------|---------------------|---------------|
| **Topic 1** | 0.021*"girl" + 0.014*"date" + 0.013*"text" + 0.011*"hang" + 0.010*"group" + 0.010*"girlfriend" + 0.009*"miss" + 0.009*"meet" + 0.008*"coupl" + 0.008*"boyfriend" | Relationship |
| **Topic 2** | 0.007*"emot" + 0.006*"matter" + 0.006*"exist" + 0.006*"believ" + 0.005*"realiz" + 0.005*"human" + 0.005*"futur" + 0.005*"real" + 0.005*"suffer" + 0.004*"sens" | Emotion |
| **Topic 3** | 0.016*"medic" + 0.016*"therapist" + 0.013*"doctor" + 0.009*"med" + 0.009*"diagnos" + 0.008*"hospit" + 0.007*"advic" + 0.006*"scar" + 0.006*"psychiatrist" + 0.006*"attack" | Clinic |
| **Topic 4** | 0.011*"job" + 0.009*"countri" + 0.008*"pay" + 0.008*"compani" + 0.007*"appli" + 0.007*"save" + 0.006*"unemploy" + 0.006*"manag" + 0.006*"quit" + 0.006*"afford" | Work |
| **Topic 5** | 0.020*"mother" + 0.015*"brother" + 0.015*"sister" + 0.012*"father" + 0.009*"abus" + 0.008*"die" + 0.007*"child" + 0.006*"young" + 0.006*"kid" + 0.006*"fight" | Family |
| **Topic 6** | 0.019*"game" + 0.016*"watch" + 0.012*"video" + 0.011*"enjoy" + 0.010*"eat" + 0.010*"drink" + 0.009*"weight" + 0.008*"food" + 0.007*"smoke" + 0.007*"walk" | Entertatiment |
| **Topic 7** | 0.032*"class" + 0.025*"studi" + 0.024*"grade" + 0.017*"fail" + 0.015*"univers" + 0.014*"student" + 0.012*"semest" + 0.010*"teacher" + 0.010*"graduat" + 0.009*"learn" | School |

We observed that for most topics in LDA(7), the top five terms are relevant to the topics ( Table 2). Most terms have ratio $\geq$ 70% of the estimated term frequency within the selected topic to the overall term frequency. Exceptions are topics # 2 (Emotion) and # 4 (Work). In topic # 2, the top 5 topic terms have the smallest overall ratio of topic term frequency to the overall term frequency, the lowest among the topics. This may indicate that most of these topic terms are not a right fit and cannot adequately represent topic #2. Similarly, in topic # 4, although the first topic term "job" thoroughly contributes to the topic (95% of occurrences), the second topic term "countri" contributes only 60% to the topic, which means this term is too common and is not suitable in this topic. The lack of exclusiveness of topic terms in topics # 2 and # 4 may be one reason why they have a relatively low class-level precisions during the multi-class classification task (Table 5, Sec. 5.3) since Machine Learning models may not be able to learn the representative and distinctive features of some topics from their corresponding terms.

## 4.2 Topic Classification

To further evaluate and understand capability of topic generation, we run classification tasks on the topics from the LDA(7). If the topics can indeed help Machine Learning models learn the features of the text data that can be used to distinguish different topic, i.e., a high validation accuracy can be achieved, we will then use them for sentiment analysis.
The LDA model assigns each document in the corpus with a list of topics it may contain. We chose the central topic which occupies the most percentage of the topic contribution in each document as the corresponding label. The resulting data sets are imbalanced (Fig 4).

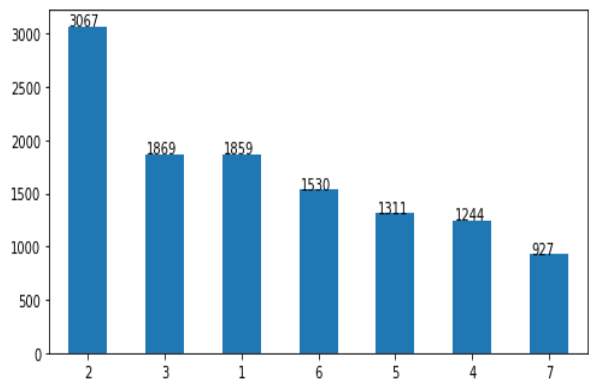

Figure 4: class distribution of 7 topics

We keep the data sets imbalanced as it has been shown that Bert can perform better on classification of imbalanced data sets than data sets that used an augmentation technique to oversample minority classes (Madabushi et al., 2020). We then perform supervised learning with the traditional models and transformer-based models, namely Naive Bayes, SVM, XLnet, and Bert. To prepare for transformer-based models for classification tasks, we fine-tuned the XLnet and Bert models from HuggingFace library (Wolf et al., 2019) with 12 layers and 768 hidden units. We used the hyperparameters of a batch size of 8, AdamW optimizer(Loshchilov & Hutter, 2019) with learning rate of 2e-5 and 4 epochs.

We used 5-fold cross validation to fully access the performance of the models. Table 3 presents the classification results.

| Machine Learning Model | Macro Precision | Macro Recall | Macro F-1 Score | Accuracy |
| --- | --- | --- | --- | --- |
| Naive Bayes | 0.459 | 0.155 | 0.082 | 0.272 |
| SVM | 0.696 | 0.651 | 0.667 | 0.673 |
| XLnet | 0.710 | 0.714 | 0.711 | 0.712 |
| Bert | 0.721 | 0.720 | 0.719 | 0.72 |

Table 3: Evaluation results for the LDA (7) topics.

The transformer-based models produce the two top results among the four models regardless of the number of topics. The best result is generated by Bert with accuracy of 72%. XLnet comes the second place with only 0.8% difference in accuracy. Further, we use the LDA(7) topics to categorize the data set for sentiment analysis.

## 5. Sentiment Analysis

### 5.1 Sentiment Assignment

Quality of sentiment labels is critical for an accurate sentiment classification of the data sets. To avoid prohibitive costs of manual labelling, we use three labelling tools: VADER (Hutto & Gilbert, 2015), which recognizes negative, positive, and neutral sentiments, Textblob (Loria, 2018), which computes the sentiment as a float from -1.0 to 1.0, and SentiWordNet (Esuli & Sebastiani, 2006), which assigns words in the sentence with corresponding positive and negative scores. Those tools identify the candidate sentiments conveyed by the text. To evaluate consistency of assigned sentiments, we ran the three tools on the entire data, i.e.,11,807 texts. Overall, the three tools agree on sentiments conveyed in 5,428 texts, Textblob and Senti – in 6,814 texts, Vader and Textblob – in 7,516 tests, and Vader and Senti – on sentiments conveyed in 8,333 texts (Fig 5).

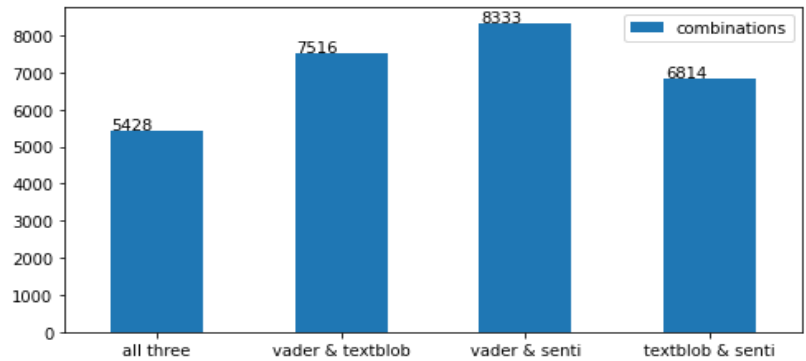

Figure 5: data count of every combination

Considering the number of posts produced by each tool combination as well as the ratio of Negative and Positive sentiments recognized by each tool (Fig 6.A) and the tools' contingency in recognition of Negative and Positive sentiments (Fig 6.B), we opt to use VAGER and Textblob jointly to determine the sentiments in our text data. This tool duo agrees on sentiments in 7,516 texts, the second biggest number of texts among all the tool combinations, and it gives the most balanced recognition of Negative and Positive sentiments, 4,662 texts and 2,854 texts, respectively. Although a slight imbalance might persist in the data set, we decide to keep the imbalance since negative sentiments entail a higher response from the Reddit users than positive sentiments (Sharma et al, 2020). We use 7,516 texts, labelled by VAGER and Tetxblob, as the benchmark data.

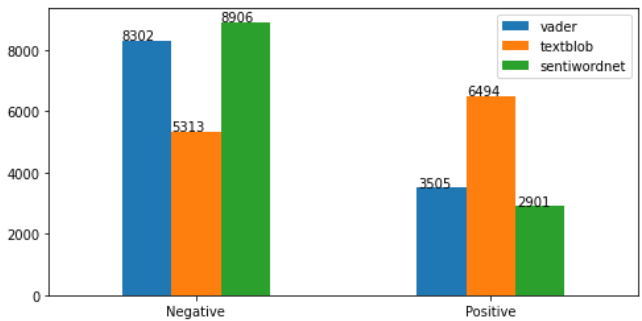

Figure 6.A: data count of N,P of different tool

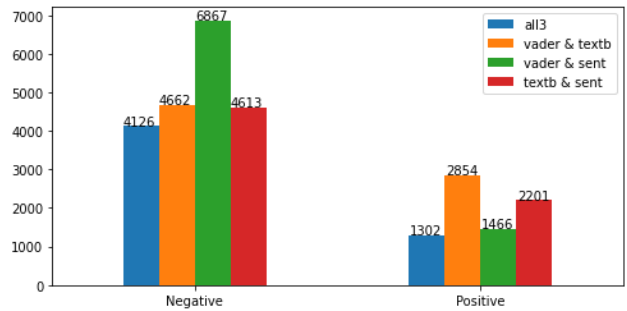

Figure 6.B: data count of N,P of different combination of tools

### 5.2 The Best Model Selection

We use 5-fold cross validation to evaluate machine learning models. We aim to choose the best model to predict sentiments of the rest of the dataset, i.e., 4,291 texts that VAGER and Textblob do not agree on their sentiments; those texts are treated as the unlabelled data. We use the hyperparameter configuration from Sec. 4.2 for the transformer-based models. Table 4 shows the evaluation statistics of the four models on the benchmark 7,516 texts.

BERT outperforms the other three models on every reported metric, with Accuracy = 91.5%. XLnet, another transformer-based model, is behind Bert by insignificant margins. Naive Bayes has the lowest performance, with Accuracy = 62.7%. Considering the running time, we opt to use BERT as our prediction model since it took significantly less time to train and proceeded with slightly better results than XLnet.    We report BERT's per class results in Table 5.

|  | Macro Precision | Macro Recall | Macro F-1 Score | Accuracy | Run time |
|---|---|---|---|---|---|
| Naive Bayes | 0.783 | 0.508 | 0.401 | 0.627 | 5.33s |
| SVM | 0.860 | 0.782 | 0.799 | 0.828 | 5.41s |
| XLnet | 0.911 | 0.901 | 0.905 | 0.912 | 3h 34min 40s |
| BERT | 0.912 | 0.907 | 0.909 | 0.915 | 2h 8min 3s |

Table 4: Sentiment classification of the fully labelled 7,516 texts.

| Class (Topic) # | Content focus | Precision | Recall | F-score |
|---|---|---|---|---|
| # 1 | relationship | 0.735 | 0.719 | 0.726 |
| # 2 | emotion | 0.718 | 0.726 | 0.722 |
| # 3 | clinic | 0.726 | 0.730 | 0.728 |
| # 4 | work | 0.713 | 0.780 | 0.744 |
| # 5 | Family | 0.727 | 0.654 | 0.688 |
| # 6 | Entertainment | 0.709 | 0.692 | 0.699 |
| # 7 | School | 0.722 | 0.739 | 0.729 |

Table 5: BERT classification results on individual classes (LDA topics)

Note that BERT can significantly outperform traditional models like Naïve Bayes, Logistic Regression, etc, in sentiment classification of online movie reviews (Gonzales-Carvajal and Carrido-Merchan, 2021). Our results extend this conclusion to sentiment classification of the Reddit posts.

## 5.3 Sentiment Analysis of the Post Topics

We fine-tuned the BERT model on the benchmark texts labelled by VADER and Textblob; there are 4,662 negative sentiments and 2,854 positive sentiments that the tools agree on. On the next step, we use the fine-tuned BERT model to assign the sentiments of the remaining 4,291 texts on which the tools disagree. The BERT model labels them with 3,070 negative sentiments and 1,221 positive sentiments. Among the 3,070 texts predicted as negatives, VADER and Textblob identified 2710 of them as negative, whereas SentiWordNet labelled 2407 texts as negative. Among the 1221 positive texts, VADER and Textblob were able to label 930 out of 1221 texts as positive. SentiWordNet was able to label 821 texts as positive.

Recall that the LDA(7) model assigns a topic to represent every text. As the result, we have 11,807 texts labelled with a sentiment and an LDA topic. Out of the 11,807 sentiment labels, we now obtain 7,464 negative labels and 4,343 positive labels.

Whereas the number of negative posts almost doubles the number of the positive posts, the positive sentiments still exist in the texts. This prompts to ask a question: Given topic X, what is the probability of its correspondence with positive and negative sentiments? Another venue is to determine which topic has a higher negativity or positivity association.

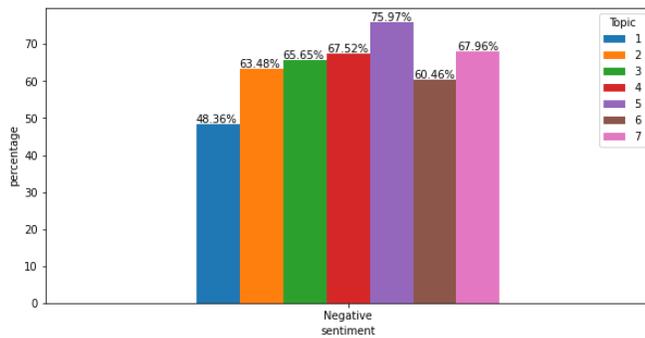
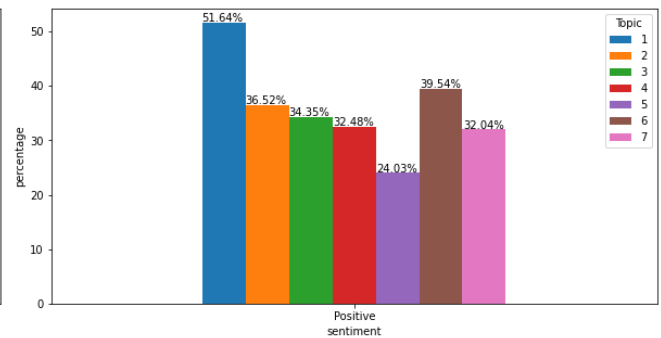

Figure 7.A: Percentage of negative sentiments given topics     Figure 7.B: Percentage of positive sentiments given topics

75.97% of the data in topic # 5 is labelled with a negative sentiment (Fig 7.A). Compared to the other topics, topic # 5 has the more negativity associated with it, whereas topic #1 has the least percentage of negativity among its assigned posts (48.36%). On the other hand, in the positive category (Fig 7.B), topic # 1 (relationship-related) has the most positive ratio, with 51.64%, leaving all other topics behind by significant margin of at least 10%, and topic# 5 has only 24.03% of positivity.

      We conclude that in 'r/depression' subreddit, family-related (topic #5) is more negative-oriented than other topics discussed under the Covid-19 situation. Our results correspond to the results by Humphreys et al. (2020) that domestic violence, especially children's abuse, increase during the quarantine and lockdown isolations and the results by Panchal et al. (2021) that

families with parents and children living together experience higher levels of depression and anxiety.

We hypothesize that the high negative rank of topic #5 is in part due to a younger demographics overrepresented on Reddit: 64 % of the Reddit user base was between the ages of 18 and 29.[7]
As the users cannot gain supports from family members, so they post and share their stories to find support and advise to help them cope with the difficulties.

### 5.4 The Study Limitations

1) When we labelled the text data with LDA topics, we only used the topic that contributed the most to the text. At the same time, a text can cover multiple topics and contributions of central topics in some texts may not significantly exceed contributions of the other topics. In this case, we suggest a) use multi-labelled topic assignment, b) use only a topic that contributes $> 40\%$ to the text to increase the confidence of labelling. With the $LDA(7)$ model, however, this will reduce the size of our data from 11,807 to 8,370 texts.
2) A more thorough analysis can be conducted with multi-class sentiment classification instead of using binary, positive and negative, classification. The shift to multi-class classification will require to use sentiment analytics tools that recognize multiple sentiments.
3) To explore the full potential of the XLnet model we want to use more than 512 characters.

## 6. Conclusions and Future Work

In this study, we have identified a set of illustrative topics that represent the COVID-related posts collected from r/Depression. We have identified what topics have a more positive connotation and what topics have a more negative connotation. In a series of empirical experiments, we have identified BERT as the best prediction model for sentiment analysis of the posts. BERT achieved its highest F-score = 0.744 in identification of the work-related topic, and its lowest F-score = 0.688 - on the family-related topic.

Compared with other extracted topics, the family-related topic has more negativity associated with it (75.97%), whereas the relationship-related topic has the least percentage of negativity among its assigned posts (48.36%). In fact, the relationship-related topic has the most positive ratio, with 51.64%, leaving all the other topics behind by significant margin of at least 10%.

Further studies can involve messages posted on other depression-related Reddit communities. Secondly, we could apply different topic modelling techniques, for example, Bayesian SMM, a state-of-the-art generative log-linear model that learns to represent corpus in the form of Gaussian distribution (Kesiraju et al., 2019). We can use Bayesian SMM to get a better topic extraction. Additionally, we could use a variant of Bert, RoBerta (Liu et al., 2019) to perform sentiment classification to make a more precise label assignment.

---

[7] https://www.alphr.com/demographics-reddit/

# Appendix: Toolkits used in the study

| Task | Name | URL |
|---|---|---|
| Reddit post extraction | PushshiftAPI | https://github.com/pushshift/api |
| Pulling data out of HTML | BeautifulSoup | https://www.crummy.com/software/BeautifulSoup/bs4/doc/ |
| Build LDA model and text processing | Gensim | https://radimrehurek.com/gensim/ |
| Contraction reconstruction | Contraction | https://github.com/kootenpv/contractions |
| Build the transformer-based model | HuggingFace | https://huggingface.co/ |
| LDA model implementation | MALLET | http://mallet.cs.umass.edu/ |
| Fine-tune transformer-based models. | AdamW Optimizer | https://huggingface.co/transformers/main_classes/optimizer_schedules.html |
| SentimenWordNet implementation and text processing | Nltk | https://www.nltk.org/ |
| Sentiment assignment | vaderSentiment | https://github.com/cjhutto/vaderSentiment |
| Sentiment analyzer | TextBlob | https://textblob.readthedocs.io/en/dev/advanced_usage.html#sentiment-analyzers |
| build traditional ML models | sklearn | https://scikit-learn.org/stable/ |